\documentclass[aps,prl,showpacs,twocolumn]{revtex4}
\usepackage{graphicx}
\usepackage{amsmath}
\usepackage{amstext}
\usepackage{latexsym}
\usepackage{amsfonts}
\usepackage{bm}
\usepackage{amssymb}
\usepackage{amscd}
\begin{document}
\title{Quantum Key Distribution with Screening and Analyzing}
\author{Won-Ho Kye}
\affiliation{The Korean Intellectual Property Office, Daejeon
302-701, Korea}
\date{\today}
\begin{abstract}
We propose a quantum key distribution scheme by 
using screening angles and analyzing detectors which enable 
to notice the presence of Eve who eavesdrops the quantum channel, as the revised protocol
of the recent quantum key distribution [Phys. Rev. Lett. 95, 040501 (2005)]. 
We discuss the security of the proposed quantum key distribution against various attacks including impersonation attack and Trojan Horse attack.

\end{abstract}
 \maketitle
Quantum key distribution (QKD) is to find a way which enable to 
share the secret information between distant parties
by transmitting quantum states. 
The security of QKD is guaranteed by the law of quantum physics, 
while that of the classical key distribution or cryptography is done 
by the computational cost of the underling mathematical problems.

Bennet-Brassard (BB84)\cite{BB84} proposed the first QKD protocol which uses 4 quantum states
and Ekert (E91) \cite{E91} suggested basing the security of the two-qubit protocol 
on Bell's inequality \cite{Gisin}.
On the other hand, Bostr\"om and Felbinger' \cite{Ping} initiates a new type of 
QKD which generates the key in deterministic way by passing the qubit in two-way.
Lucamarini and Mancini's protocol \cite{Luca} is also classified in this type QKD.
The deterministic feature of the two-way protocols enables to apply the direct encoding but
in some case it was considered as a vulnerable point.  

The situation in regard to the trend of QKD looks contradictory in the sense that
QKD should be implemented with single photon or sufficiently 
weak pulse to guarantee the security, even though the coding should be done with 
strong pulse to send the signal to longer distance.
In this regard, Kye et al. \cite{Kye} has proposed 3-way type QKD protocol which 
enables to send not-so-weak coherent pulses 
by loading the encoding onto the qubit with 
perfect random polarization. 
Since the first version of the protocol is vulnerable against 
the impersonation attack when it is implemented with single pulse,
they have proposed the protocol using two coherent pulses.
Unfortunately, there has been several attacks \cite{att1, att2, Kim} and
the revised protocols \cite{Kim, sol1, sol2} are followed.
Even though the revised version of the protocol is secure in present, 
the fact that two coherent pulses are used to 
encode qubit has been considered as the attack
point \cite{att1, att2, sol1, sol2}.

In this paper, keeping the basic philosophy of QKD in Ref. \cite{Kye},
we propose a new scheme which blocks up all known attacks \cite{att1, att2} and 
the composite attack which is composed of PNS and Trojan Horse attack.
The proposed QKD protocol is described as follows:

\begin{enumerate}

\item[(P.1)] Alice announces a set $S(N)$ which has $N$ number of screening angles,  
\begin{eqnarray}
 S(N)= \{ \alpha_1, \cdots \alpha_N \}, 
\end{eqnarray}
where the screening angle $\alpha_i$ is defined as $\alpha_i=i\pi/2(N+1)$.
Alice prepares the analyzing detectors (AD) in her setup (see Fig. 1).
 
\item[(P.2)] Alice
prepares a qubit with arbitrary angle $\theta$.
\begin{equation}
	|\theta\rangle,
\end{equation}
She sends the qubit to Bob.

\item [(P.3)] Bob choose a random angle $\phi$ and a screening angle $\alpha_b$. 
He acts $U(\phi+\alpha_b)$ on the received qubit.
The qubit becomes
\begin{equation}
|\theta+\phi+\alpha_b\rangle,
\end{equation}
where Bob occasionally takes $\phi$ as an analyzing angle
$\phi^* \in \{0, \pi/2\}$ with the probability $p_a$.
Bob sends the qubit to Alice.

\item [(P.4)] After receiving the qubit, Alice applies  $U(-\theta+(-1)^k\pi/4 +\alpha_a)$, where $\alpha_a$ is a chosen screening angle. The qubit becomes 
\begin{equation}
|\phi + (-1)^k \pi/4 + \alpha_a+\alpha_b\rangle,
\end{equation}
The fraction $(1-t)$ of the photons in the qubit clicks the Alice's AD $D_1$ and $D_2$.
If Bob takes $\phi=\phi^*$ and screening angles $\alpha_a$ and $\alpha_b$
are matched as
\begin{equation}
\alpha_a+\alpha_b=\pi/2,
\label{eq.sum}
\end{equation}
then the qubit incoming to Alice's detector is $|\phi^* + (-1)^k \pi/4 + \pi/2\rangle$. 
Thus the outcome on the Alice's detector $O_a$ and Alice's encoded key $k$ are related by 
\begin{equation}
O_a=k\oplus(2\phi^*/\pi)\oplus 1.
\label{eq.oa}
\end{equation}
The above equation shows Alice's integrity condition in AD.
The remaining fraction $t$ of the photons in qubit is sent to Bob.

\item[(P.5)] Bob 
applies $U(-\phi)$ on the received qubit.
The qubit becomes
$|(-1)^k \pi/4 + \alpha_a+\alpha_b\rangle$.
Bob measures the received qubit in $(+\pi/4, -\pi/4)$ basis and get the outcome  $O_b$.
The Bob's outcome is definitely correlated with the encoded key $k$ only 
when the screening angles are matched as Eq. (5)
and the key $k$ and Bob's outcome $O_b$ are related by
\begin{equation}
O_b= k \oplus 1.
\end{equation}

\item [(P.6)] After repeating $M$ times from (P.2) to (P.5),
Alice and Bob disclose the series of screening angles $\alpha_a$ and $\alpha_b$ and the
sequence of analyzing angle with $\phi^*$ and the value of $\phi^*$. 
Alice and Bob create the key $k_a$ and $k_b$ only with the sequence satisfying Eq. (\ref{eq.sum}) and
exchange the hash values $h(k_a)$ and $h(k_b)$ of the created $k_a$ and $k_b$ \cite{Kye}.
If $h(k_a)=h(k_b)$ and Eq. (\ref{eq.oa}) is verified for all analyzing angles $\phi^*$, 
key creation is finished else Alice and Bob start again from (P.1). 
\end{enumerate}

\begin{figure}
\rotatebox[origin=c]{0}{\includegraphics[width=7.5cm]{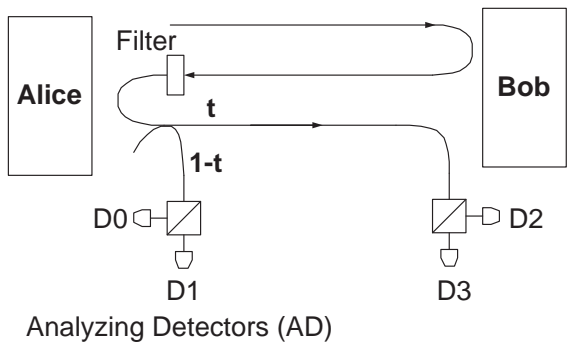}}
\caption{Setup of the proposed QKD.}
\label{figure}
\end{figure}
In general, the probability of the event which is satisfied with Eq. (6) is $N/N^2=1/N$
thus the key creation rate is proportional to $1/N$ in this QKD.
For example, the simplest implementation is plausible with $N=2$. 
In that case, $S(2)=\{\pi/6, 2\pi/6\}$ and 
$50\%$ of the transmitted quits are satisfied with the matching condition Eq. (\ref{eq.sum}).
So Alice and Bob can utilize $50\%$ of transmitted quits in creation of key. 
The screening angles and AD play an 
important role to detect the presence of Eve.
As we shall see, the screening angle unable Eve to relay the qubit 
without inducing the measurement error in Bob and AD blocks up 
general Trojan Horse attack and more sophisticated one combined with PNS.

Since Alice and Bob only utilize the qubit event satisfying Eq. (6) in creation of the key,
this QKD is not deterministic protocol. That is neither Alice nor Bob expect the created key 
and it is determined only after the announcement of screening angles, which is the
original concept of BB84 \cite{BB84}.  
This is different feature compared with the original QKD in Ref. \cite{Kye}.
Here we shall show that the proposed QKD with screening angle and AD
is secure against the known attacks \cite{att1, att2}\\

{\it Security against Impersonation attack}:\\
In regard to the impersonation attack, we shall consider the case that Alice turns off 
AD by setting the transmission coefficient $t=1$ for the demonstration 
of the role of screening angles clearly. 
 
\begin{enumerate}
\item[(1)] Single photon Implementation:
\begin{enumerate}
\item [(A.1)]
After (P.2), Eve intercepts and stores the pulse from
Alice in set $E_1$. Thus Eve has $E_1=\{|\theta\rangle\}$. 
Eve sends to Bob a pulse originally
prepared by her with random angles $\theta^\prime$.

\item [(A.2)]
After step (P.3), Eve intercepts both pulses from Bob and stores
them in set $E_2$ after compensating with the angles
$-\theta^\prime$. Eve then has $E_2=\{|\phi+\alpha_b\rangle\}$. 
Eve send $E_1$ qubit to Alice.

\item [(A.3)]
After step (P.4), Eve intercepts the returning qubit
$|(-1)^k\pi/4+\alpha_a\rangle$.
Eve need to guess $\alpha_a$ to measure the qubit without error.
If Eve's guessing is correct, Eve can read Alice's encoding and 
Eve get the outcome of the measurement $O_e=k$.
Eve encode the $O_e$ into $E_2$ qubit then the qubit becomes 
$E_2^\prime=\{|\phi + (-1)^{k}\pi/4 +\alpha_b\rangle \}$.
Eve send the qubit to Bob.

\item[(A.4)] 
After the step (P.5), the receive qubit becomes $E_2^{\prime\prime}=\{|(-1)^{k}\pi/4 + \alpha_b \rangle \}$
Thus the Bob measurement includes the error induced by Eve's impersonation as follows: 
\begin{equation*}
\mbox{IE}(N)\sim \sum_{i=1}^N\sin^2(\alpha_i-\pi/2),
\end{equation*}
which is the lowest bound of the error due to Eve's impersonation.
Thus it should be noticed in the key verification stage of (P.6).
If Eve's guessing was not correct, Eve reads incorrect encoding and the error is additionally 
propagated into the $E_2^\prime$ and Bob's key. It also should be noticed in the stage of (P.6).

\end{enumerate}

\item[(2)] Pulse implementation:

In this case, Eve's attack strategy is different from that of
the single photon implementation.
Eve can use beam splitter to separate the pulse into two parts and
Eve tries to analyze the pulse measuring it on both bases  
$(\alpha_1+\pi/4, \alpha_1-\pi/4)$ and $(\alpha_2+\pi/4, \alpha_2-\pi/4)$.
However in this experiment Eve can not get the conclusive results 
because two bases are not orthogonal with each other. Thus it eventually 
induces the error in Bob's key as a results of Eve's relay 
based on inconclusive result.

If Alice turns on the AD with $t <1$, Eve's impersonation without knowing
the $\phi$ should violate the Alice's integrity condition Eq. (\ref{eq.oa}). 
So Alice should notice the presence of Eve more easily.
\end{enumerate}

{\it Security against photon number splitting (PNS) attack}:\\
Since Alice and Bob announce the screening angles in (P.6), 
the security against PNS attack determined by the 
two random angle $\theta$ and $\phi$.
Thus the security is 
almost same with that of single pulse implementation
of QKD in Ref. \cite{Kye}.
The only difference is that the random 
angle $\phi$ is partially known to Eve 
when Bob chooses the analyzing angle $\phi=\phi^*$ .\\

{\it Security against Trojan Horse type attack}:\\
To eavesdrop the key, Eve needs to distinguish the her 
injected photon from those in the pulse 
and separate it out correctly. Thus Eve necessarily marks the 
injected photon by using time delay or frequency shift \cite{Trojan}.
However it is known that this type of marking
can be blocked with properly designed apparatuses and filters \cite{Trojan}. 
Accordingly, we assume that Alice has a properly designed filter 
as in Fig. 1.

It is clear that the proposed QKD is robust against simple Trojan Horse attack,
because in (P.4) Alice's encoding is performed with the random angle 
compensation of $\theta$ such that $U(-\theta+(-1)^k\pi/4+ \alpha_a)$.
Accordingly, the information gain for Eve, who attacks quantum channel
with independent Trojan Horse state $|\eta\rangle$, 
is zero because the separated qubit always has the random polarization.
Thus the general Trojan Horse attacks discussed in Ref. \cite{Trojan}
are not useful for Eve in our QKD.

However Eve can consider the following strategy which 
combines PNS and Trojan Horse attack.
\begin{enumerate}
\item[(B.1)] After (P.2) Eve performs quantum quit measurement 
on the pulse and if the number of
photons are larger than one, Eve separate out
a photon and keeps it in $E_1=\{|\theta\rangle\}$.
\item[(B.2)]Eve attaches the photon $|\theta\rangle$ 
onto the incoming pulse after (P.3).
The state of incoming photon pulse is given by $|\theta + \phi +\alpha_b\rangle \otimes |\theta\rangle$.
After (P.4) the outgoing pulse is given by 
\begin{equation}
|(-1)^k \pi/4+\phi+\alpha_a+\alpha_b\rangle \otimes |(-1)^k \pi/4 +\alpha_a\rangle.
\end{equation}
\item[(B.3)]Eve keeps $E_2 =\{|(-1)^k\pi/4 +\alpha_a\rangle \}$ after separating the photon from the pulse. Eve reads the key value $k$ by measuring $E_2$ 
after the announce of $\alpha_a$ in (P.6).
\end{enumerate}
With this strategy, if Eve separates out the $|(-1)^k \pi/4 +\alpha_a\rangle$, 
she can eavesdrop the $k$ after the announcing of $\alpha_a$ in (P.6) 
by measuring the separted qubit \cite{Eve}. 

We can see that this strategy of Eve is easily noticed by the help of AD.
The fraction $(1-t)$ of photons in the qubit in Eq. (8) clicks on
Alice's AD. The mixed state $|(-1)^k \pi/4+\phi+\alpha_a+\alpha_b\rangle \otimes |(-1)^k \pi/4 +\alpha_a\rangle$ eventually give rise to
the violation of integrity condition in Eq. (\ref{eq.oa}) \cite{Integrity}, 
because the equation holds only when the pure state 
$|(-1)^k \pi/4+\phi+\alpha_a+\alpha_b\rangle$
is entered.

Eve may consider the attack proposed in Ref. \cite{Cai} which attaches the
standard state $|0\rangle$ instead of PNS photon $|\theta\rangle$ 
in (B.2) and calculates the polarization directly 
estimating maximal mean fidelity.
The corresponding qubit is
\begin{equation}
|(-1)^k \pi/4+\phi+\alpha_a+\alpha_b\rangle \otimes |-\theta+(-1)^k \pi/4 +\alpha_a\rangle.
\end{equation}
This qubit also violate the Alice's integrity condition in AD (Eq. (\ref{eq.oa}))  \cite{Integrity}.

By announcing the larger number of screening angles $N$, Alice and Bob
may get rid of possibility of the leakage of the information encoded on the qubit.
To attack the protocol which uses the larger number of screening angle $N$,
Eve need to prepare $N$ separated pulse by using the beam splitter
and to measure the pulses with $N$ different nonorthogonal basis.
Accordingly, the probability of which Eve get the conclusive result for relay is suppressed depending on $N$
thus we can say that the security level of this scheme is proportional to $N$, while the 
key creation rate is  proportional to $1/N$. We note the Bob's choice of the analyzing angle $\phi^*$ does not affect the key creation rate because $\phi^*$ is also one of the random angle that Bob can choose in (P.3).
Alice and Bob establish the desired security level by controlling the
number of screening angles $N$, appropriately.
It can be summarized as follows:
\begin{eqnarray}
\mbox{Key Creation Rate} &\sim& 1/N, \\
\mbox{Security Level} &\sim& N.
\end{eqnarray}

In conclusion, we have proposed the revised QKD protocol
which enables to implement QKD with not-so-weak single pulse.  
We have shown that the screening angles and AD enable Alice and Bob not only to notice 
the impersonation attack but also to block up the
Trojan Horse attack and the composite attack with PNS.
The proposed QKD is simple to implement and provides tunable security level depending $N$.



\begin{thebibliography}{150}
\bibitem{Integrity} In AD, the events which click $D1$ and $D2$ simultaneously  are also considered as the violation of integrity condition, 
because the value of $O_a$  in Eq. (\ref{eq.oa}) is inconclusive.
\bibitem{Eve} In this stage, we assume that Eve has identified injected 
photon
as well as separted it successfully amoung the identical photons 
inside of coherent pulse by her unlimited
technology, even though this is not the 
technical problem but principle one.
\end{thebibliography}
\end{document}